# "Graphene-Like" Exfoliation of Atomically-Thin Bismuth Telluride Films

**Desalegne Teweldebrhan, Vivek Goyal, Muhammad Rahman and Alexander A. Balandin**[∗]

*Nano-Device Laboratory*
*Department of Electrical Engineering and Materials Science and Engineering Program*
*Bourns College of Engineering*
*University of California – Riverside, Riverside, California 92521 USA*

## Abstract

We report on "graphene-like" exfoliation of the large-area crystalline films and ribbons of bismuth telluride with the thicknesses of a few atoms. It is demonstrated that $Bi_2Te_3$ crystal can be mechanically separated into its building blocks – Te-Bi-Te-Bi-Te *atomic five-folds* – with the thickness of ~1 nm and even further – to subunits with smaller thicknesses. The atomically-thin films can be structured into suspended crystalline ribbons providing quantum confinement in two dimensions. The quasi two-dimensional crystals of bismuth telluride revealed high electrical conductivity. The proposed *atomic-layer engineering* of bismuth telluride opens up a principally new route for drastic enhancement of the thermoelectric figure of merit.

---

[∗] Corresponding author; electronic address: balandin@ee.ucr.edu ; group web-site: http://ndl.ee.ucr.edu

D. Teweldebrhan, V. Goyal, M. Rahman and A.A. Balandin, University of California - Riverside, 2009

The first mechanical exfoliation of graphene [1] and discovery of its unique electrical [2-3], thermal [4-5] and optical [6] properties stimulated major interest to the atomically-thin films. In this letter we show that bismuth-telluride ($Bi_2Te_3$), a vital material for thermoelectric industry, can also be subjected to mechanical exfoliation resulting in the atomically-thin crystalline films, i.e. quasi two-dimensional (2-D) crystals. The successful exfoliation of the quasi 2-D crystals of bismuth telluride opens up tremendous opportunities for re-engineering its thermoelectric properties and enhancing the thermoelectric figure of merit $ZT=S^2\sigma T/K$ (here $S=-\Delta V/\Delta T$ is the Seebeck coefficient, $\Delta V$ is the voltage difference corresponding to a given temperature difference $\Delta T$, $\sigma$ is the electrical conductivity and $K$ is the thermal conductivity). It has been predicted theoretically that $ZT$ can be drastically increased in crystalline $Bi_2Te_3$ quantum wells with the thickness of just few atomic layers (~1 nm) owing to either charge carrier confinement [7-9] or acoustic phonon confinement [10-11]. Conventional chemical vapor deposition, electrochemical or other means are not suitable for fabrication of crystalline structures with such a thickness. Most of the thermoelectric thin films or superlattices on the basis of $Bi_2Te_3$ investigated so far were either polycrystalline or alloyed, or had the thicknesses far greater and the potential barrier higher far less than those required for strong spatial confinement of electrons and phonons.

In this letter, we show that quasi 2-D crystals of bismuth telluride can be mechanically exfoliated following a "graphene-like" procedure. The presence of the van der Waals gap allows one to disassemble $Bi_2Te_3$ crystal into its building blocks – five mono-atomic sheets of $Te^{(1)}$-Bi-$Te^{(2)}$-Bi-$Te^{(1)}$, which have the thickness of ~1 nm. In some cases, the atomic five-folds can be broken further leading to Bi-Te atomic bi-layers and Bi-Te-Bi atomic tri-layers. The resulting quasi-2-D crystals retain their good electrical conduction important for thermoelectric applications. The crystal structure of $Bi_2Te_3$ is rhombohedral with five atoms in one unit cell [12]. The lattice parameters of the hexagonal cells of $Bi_2Te_3$ are $a_H$= 0.4384 nm and $c_H$ = 3.045 nm. Its atomic arrangement can be visualized in terms of the layered sandwich structure (see Figure 1). Each sandwich is built up by five mono-atomic sheets {-$Te^{(1)}$-Bi-$Te^{(2)}$-Bi-$Te^{(1)}$-}, referred to as atomic five-folds, along the $c_H$ axis. The superscripts (1) and (2) denote two different chemical states for the anions. The outmost atoms $Te^{(1)}$ are strongly bound to three planar $Te^{(1)}$ and three Bi metal atoms of the same five-fold (also referred to as *quintuple*) layers and weakly bound to three $Te^{(1)}$

D. Teweldebrhan, V. Goyal, M. Rahman and A.A. Balandin, University of California - Riverside, 2009

atoms of the next five-fold. The presence of these van der Waals bonds between the five-folds allows one to break them and exfoliate the five-atomic-units rather easily. The bond strength within the five-fold are not the same. The Bi-Te$^{(2)}$ bond is the second weakest points within the crystal structure. The latter creates a possibility for producing quasi 2-D atomic tri-layers and bi-layers.

In order to isolate Bi$_2$Te$_3$ five-folds and break them into atomic planes we employed a method similar to the one used for exfoliation of single-layer graphene [1-3]. We have previously had extensive experience with mechanical exfoliation of graphene from various bulk graphite sources [13] and determining the number of layers through Raman spectroscopy on different substrates [14]. In the present case of bismuth telluride, the number of atomic planes was verified through a combination of the optical inspection, atomic force microscopy (AFM) and the scanning electron microscopy (SEM). The thickness of the atomic five-fold is ~1 nm. We scanned the step-like edges of the flakes with an advanced AFM (Veeco Dimension ICON) capable of ~0.1 nm vertical resolution. Some of the obtained few-atomic layer flakes were transferred to the Si/SO$_2$ wafers with prefabricated trenches for better visualization (see SEM images in Figure 2 (a)). We found that the Bi-Te atomic planes can also be broken into long ribbons as the one shown in Figure 2 (b). The latter can potentially lead to the electron (hole) quantum confinement in two lateral dimensions owing to the very small effective masses of the charge carriers. The rectangular and rhomb-shaped films were present among the exfoliated flakes, which facilitated the device fabrication. Based on AFM and SEM inspection the rhomb-shaped crystalline film shown in Figure 2 (c) is an atomic {-Te$^{(1)}$-Bi-Te$^{(2)}$-Bi-Te$^{(1)}$-} five-fold. It is interesting to note that it thickness of ~1 nm should be sufficient for a drastic enhancement of *ZT* in the bismuth telluride quantum well according to the predictions of Dresselhaus *et al.* [7]. Note that in our mechanically exfoliated Bi-Te quasi 2-D crystals the potential barrier height is essentially infinite unlike in the Bi$_2$Te$_3$-based superlattices. The latter creates interesting possibilities for the electron and phonon band-structure engineering, which were not available previously.

Another important feature of our quasi 2-D bismuth telluride atomic films is their perfect crystalline nature, which was verified by the selected area electron diffraction (SAED). We used SAED crystallographic technique because it is more sensitive to the in-plane atomic arrangement

D. Teweldebrhan, V. Goyal, M. Rahman and A.A. Balandin, University of California - Riverside, 2009

as compared to other techniques such as x-ray diffraction (XRD). Many spots in the atomic films have been examined with this technique. The pattern shown in Figure 3 (a) is characteristic for the hexagonally structured sub-lattices of the rhombohedral crystal lattice of $Bi_2Te_3$. In order to test the electrical properties of the obtained atomically-thin films we fabricated device structures with Ti/Au contacts on the top surface. The thickness of the $SiO_2$ layer was chosen to be ~300 nm (i.e. the same as in graphene devices) for better optical identification. A representative device is shown in Figure 3 (b). The Bi-Te flake in the middle is the atomic five-fold. The contacts seen on the image are to provide source – drain (SD) voltage. The gate bias was supplied from the back through the heavily doped Si wafer.

The room-temperature (RT) SD current – voltage characteristics of the devices made from the bismuth telluride flakes with the uniform thickness (number of atomic planes) revealed linear dependence (see inset to Figure 4) and a rather low electrical resistivity on the order of ~$10^{-4}$ $\Omega$m. The measured RT resistivity is comparable to the values reported for thick $Bi_2Te_3$ films used in thermoelectric devices [15-19]. The resistivity of ~$10^{-5}$ $\Omega$m is considered to be optimal in conventional $Bi_2Te_3$ films because its further reduction leads to decreasing Seebeck coefficient. A weak non-linearity was observed for higher SD voltage and higher temperature $T$. Figure 4 shows $I_{SD}$ current in the atomically-thin crystalline bismuth telluride films as the function of temperature for several different voltages. One can see that the current is nearly constant for $T$ below ~375 – 400 K but then starts to decrease rapidly before it saturates at T ~ 450 K. Such temperature dependence was reproducible for several tested devices and observed before and after annealing. The data shown in the inset for a different device revealed a similar trend. In general, the increase of the resistance with increasing temperature is more typical of metals rather than semiconductors. But it is rather common for $Bi_2Te_3$ films and was observed for materials produced by a range of different techniques [15-19]. It is explained by the specifics of the electron scattering on acoustic phonons and defects in $Bi_2Te_3$ [20] although few exceptions from this dependence were also reported [21]. In our case, the dependence is not monotonic with the bending point ~400 K. In the thin films with the thickness of just few atomic layers the electron transport may strongly depend on the coupling to the substrate and remote impurity scattering.

In conclusion, we succeeded in the "graphene-like" exfoliation of the crystalline films of bismuth telluride films with the thicknesses of few atoms. It was established from the comprehensive microscopic study that $Bi_2Te_3$ crystal most readily cleaves into its building blocks, {-$Te^{(1)}$-Bi-$Te^{(2)}$-Bi-$Te^{(1)}$-} atomic five-folds (also referred to as quintuples), although the tri-layer and bi-layer subunits with the smaller thicknesses were also observed. The atomic five-folds of bismuth telluride reveal high electrical conductivity with unusual temperature dependence. The obtained results open up a possibility of the atomic-layer engineering of bismuth telluride properties and may lead to a drastic enhancement of the thermoelectric figure of merit. In addition to providing a new route for engineering properties of thermoelectric materials via the electron and phonon confinement, our mechanically exfoliated quasi 2-D bismuth telluride crystals are ideal for the experimental verification of recent ideas for increasing ZT via perpendicularly applied electric field [22]. The developed exfoliation technique can also be extended to other thermoelectric materials [23].

*Acknowledgements*

The authors acknowledge the support from DARPA – SRC through the FCRP Center on Functional Engineered Nano Architectonics (FENA) and Interconnect Focus Center (IFC) as well as from US AFOSR through contract A9550-08-1-0100.

**References**

[1] Novoselov, K. S.; Geim, A. K.; Morozov, S. V.; Jiang, D.; Zhang, Y.; Dubonos, S.V.; Grigorieva, I. V.; and Firsov A. A. *Science*, **2004**, *306*, 666.

[2] Novoselov, K.S., Geim, A.K., Morozov, S.V., Jiang, D., Katsnelson, M.I., Grigorieva, I.V., Dubonos. S.V., and Firsov A. A. *Nature* **2005**, *438*, 197-200.

[3] Zhang, Y., Tan, J. W., Stormer, H. L. & Kim, P. *Nature* **2005**, *438*, 201–204.

[4] Balandin, A.A.; Ghosh, S.; Bao, W.; Calizo, I.; Teweldebrhan, D.; Miao, F. and Lau, C.N.; *Nano Letters* **2008**, *8*, 902.; Ghosh, S.; Calizo, I.; Teweldebrhan, D.; Pokatilov, E.P.; Nika, D.L., Balandin, A.A.; Bao, W.; Miao, F.; and Lau, C.N. *Applied Physics Letters* **2008**, *92*, 151911.

[5] Nika, D.L.; Ghosh, S.; Pokatilov, E.P.; Balandin, A.A. *Appl. Phys. Lett*. **2009,** *94*, 203103; Nika, D.L.; Pokatilov, E.P.; Askerov, A.S.; and Balandin, A.A. *Physical Review B* **2009**, *79*, 155413.

[6] Nair, R.R., Blake, P., Grigorenko, A.N., Novoselov, K.S., Booth, T.J., Stauber, T., Peres, N.M.R., and Geim, A.K. *Science* **320**, 1308 (2008).

[7] Dresselhaus, M.S., et al. *Physics of the Solid State* **1999**, 41, 679.

[8] Hicks, L. D.; Dresselhaus, M. S. *Phys. Rev. B.* **1993**, *47*, 12727.

[9] DiSalvo *F.J. Science* **1999**, *285*, 703.

[10] Balandin, A.;Wang, K.L. *Physical Review B* **1998**, *58*, 1544.

[11] Balandin A. and Wang K.L. *J. Applied Physics* **1998**, *84*, 6149.

[12] Richter, W.; Kohler, H.; Becker, C. R. *Phys. Stat. Sol. B,* **1977**, *84*, 619.

[13] Calizo, I.; Balandin, A.A.; Bao, W.; Miao, F.; and Lau, C. N. *Nano Letters* **2007**, *7*, 2645; Calizo, I.; Miao, F.; Bao, W.; Lau, C.N.; and Balandin, A.A. *Applied Physics Letters* **2007**, *91*, 071913.

[14] Calizo, I.; Bao, W.; Miao, F.; Lau, C. N.; and Balandin A.A. *Applied Physics Letters* **2007**, *91*, 201904; Calizo, I.; Bejenari, I; Rahman, M.; Liu, G. and Balandin, A.A. *J. Applied Physics* **2009**, *106*, 043509.

[15] Ji, X.H., Zhao, X.B.; Zhang, Y.H.; Lu, B.H.; Ni, H.L. *Materials Letters*, **2005**, *59*, 682.

[16] Yamashita, O.; Tomiyoshi, S. *Jpn. J. Appl. Phys*. **2003**, *42*, 492-500.

[17] Xie, W.; et al. *J. Appl. Phys*. **2009**, *105*, 113713.

[18] Fleurial, J.P.; Gailliard, L.; Triboulet, R. *J. Phys. Chem. Solids*. **1988**, *49*, 1237 -1247.

[19] Kulbachinskii, V.A., et al. *Phys. Rev. B*. **1994**, *50*, 16921.

[20] Christakudi, T.A.; Plachkova, S.K.; Christakudis, G. C. *Phys. Stat. Sol. (b).* **1996**, *195*, 217.

[21] Damodara Das, V.; Soundararajan, N.; *Phys. Rev. B*, **1988**, *37*, 4552.

[22] Bejenari, I.; Kantser, V. *Phys. Rev. B* **2008**, 78, 115322.

[23] Casian, A.; Sur, I.; Scherrer, H.; Dashevsky, Z. *Phys. Rev. B* **2000**, 61, 15965.

D. Teweldebrhan, V. Goyal, M. Rahman and A.A. Balandin, University of California - Riverside, 2009

**FIGURE CAPTIONS**

**Figure 1:** Schematic of {-Te$^{(1)}$-Bi-Te$^{(2)}$-Bi-Te$^{(1)}$-} atomic five-fold as referred to as quintuples, which are building blocks of Bi$_2$Te$_3$ crystal. The five-folds are bound to each other via weak van der Waals forces, which allow for their mechanical exfoliation. The blue, green and red colors correspond to Te$^{(1)}$, Bi and Te$^{(2)}$ atoms.

**Figure 2**: SEM images of quasi-2-D bismuth telluride crystals demonstrating the possibilities of the exfoliation technique. (a) Few-atomic-layer Bi-Te crystals suspended across a trench in Si/SiO$_2$ wafer. Note that the sides of the trench are clearly seen through the film owing to its few-atomic thickness. (b) Long (> 6 μm) suspended ribbon made of quasi-2-D bismuth telluride crystal. (c) Large-area uniform atomic five-fold of rhombic shape.

**Figure 3**: (a) SAED pattern indicating that the atomically thin films of bismuth telluride are crystalline. (b) Fabricated device structure to test the electrical properties of the quasi-2-D bismuth telluride crystals.

**Figure 4:** Current as a function of temperature in Bi-Te atomic crystals shown for different source – drain voltages. Inset shows current – voltage characteristics in the low-bias region for different temperature.

D. Teweldebrhan, V. Goyal, M. Rahman and A.A. Balandin, University of California - Riverside, 2009

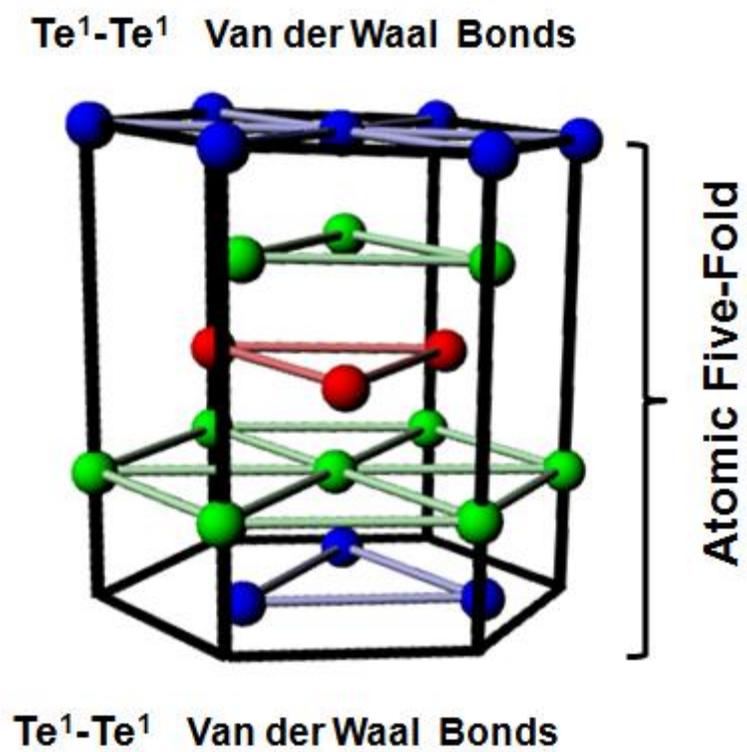

Figure 1

D. Teweldebrhan, V. Goyal, M. Rahman and A.A. Balandin, University of California - Riverside, 2009

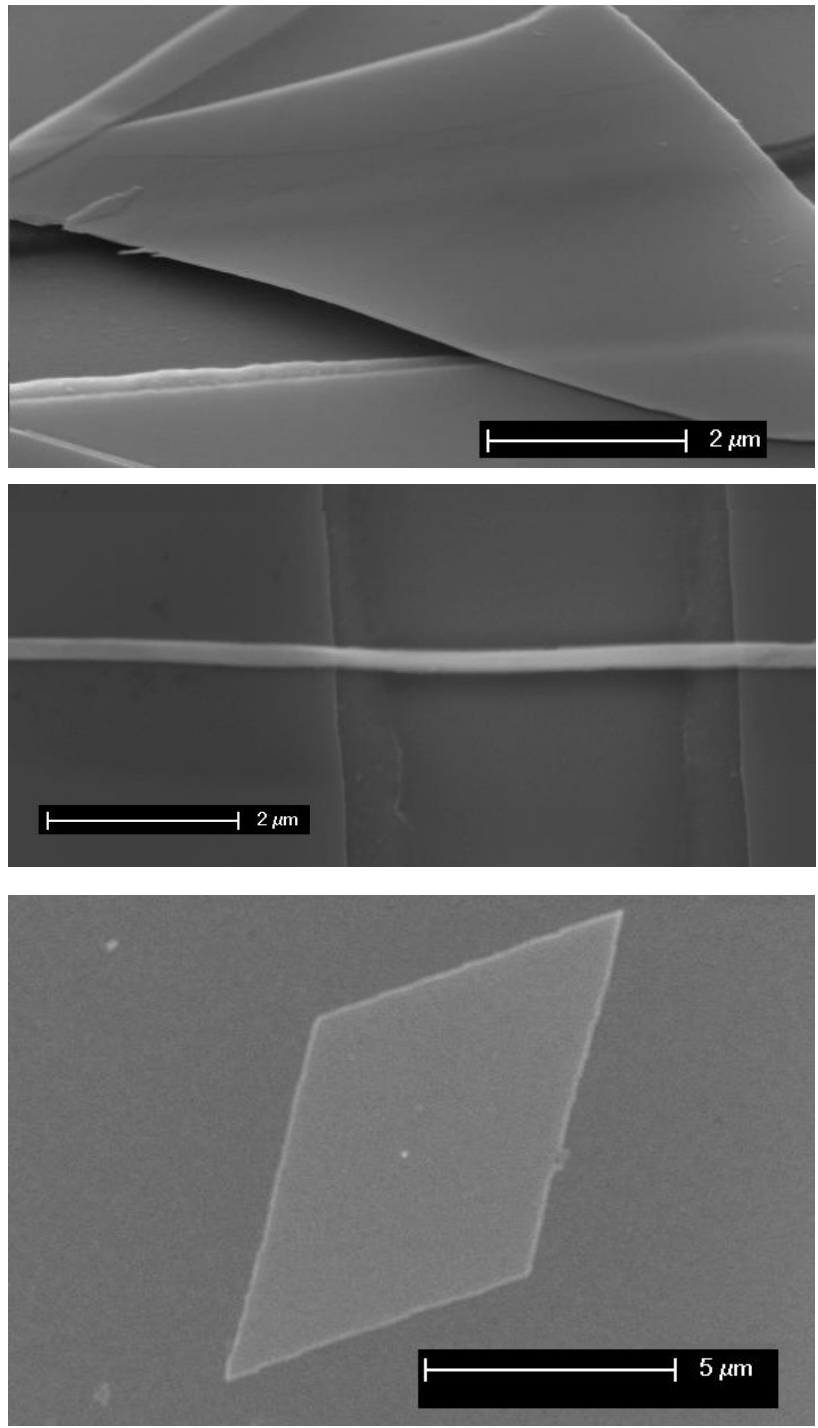

Figure 2 (A-C)

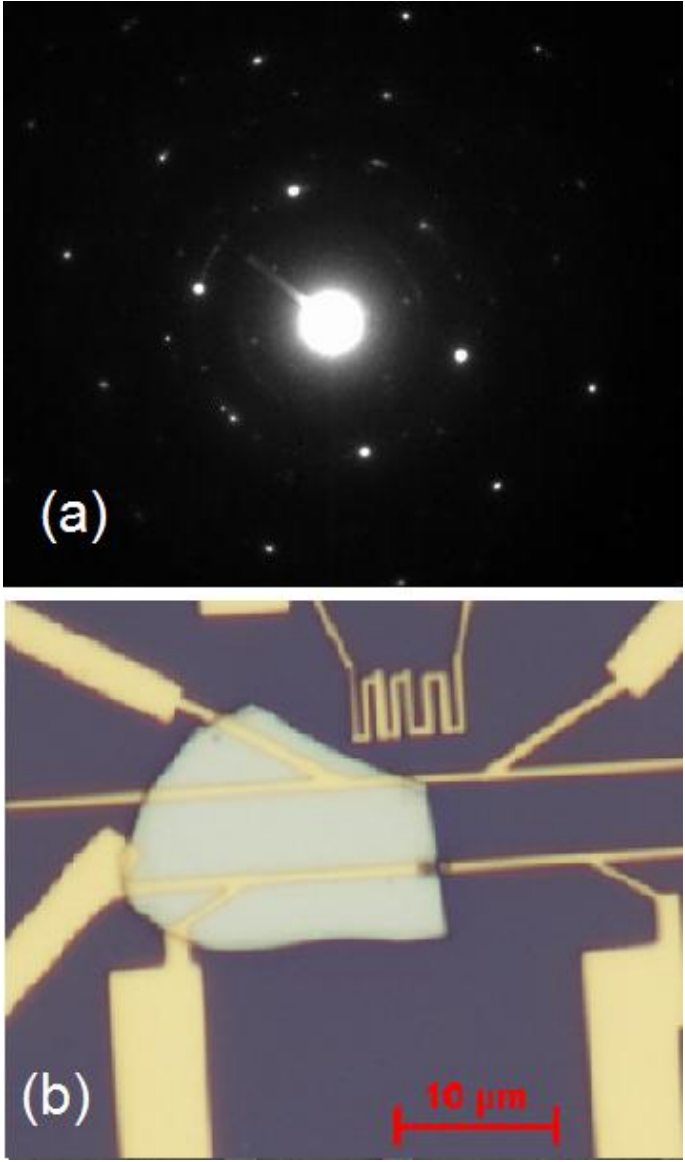

Figure 3

D. Teweldebrhan, V. Goyal, M. Rahman and A.A. Balandin, University of California - Riverside, 2009

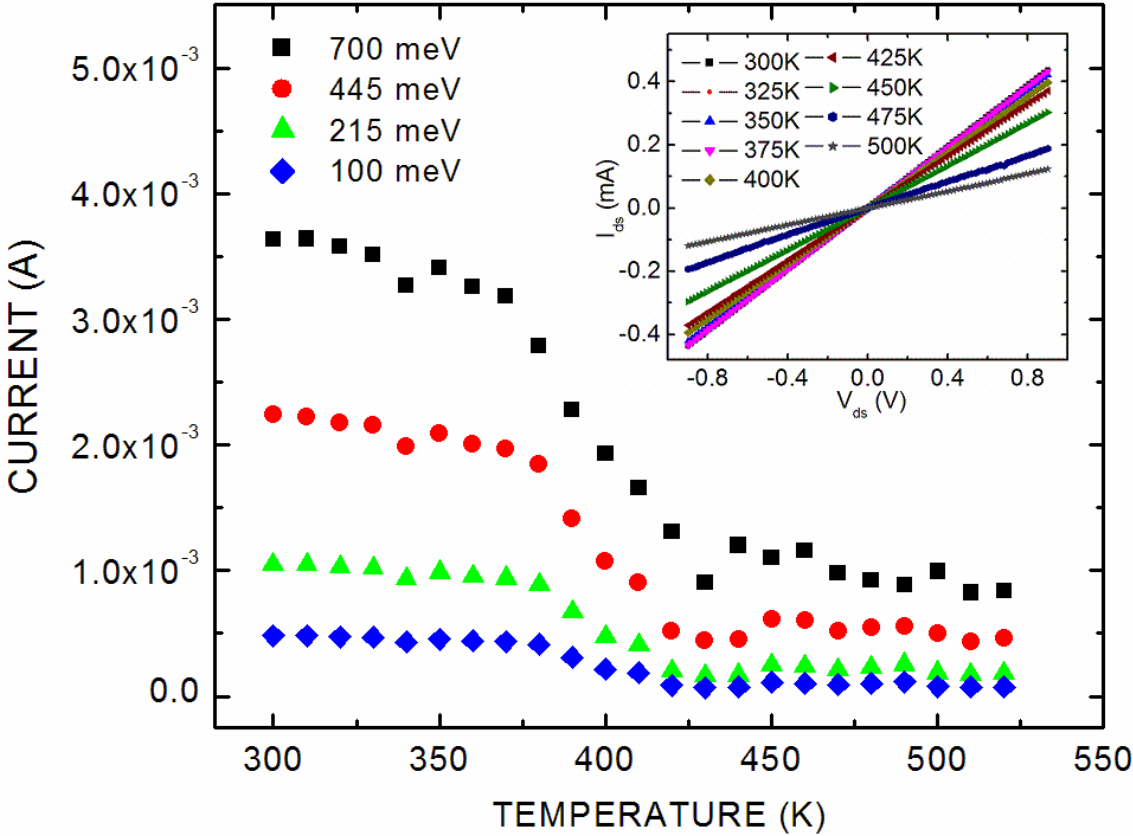

Figure 4